\documentclass[fleqn,usenatbib]{mnras}

\usepackage{newtxtext,newtxmath}
\usepackage[T1]{fontenc}

\DeclareRobustCommand{\VAN}[3]{#2}
\let\VANthebibliography\thebibliography
\def\thebibliography{\DeclareRobustCommand{\VAN}[3]{##3}\VANthebibliography}

\usepackage{graphicx}	% Including figure files
\usepackage{subfig}
\usepackage{amsmath}	% Advanced maths commands
\title[]{Identification of orientation of galaxies in the Galaxy Zoo dataset using spectral clustering }

\author[]{
Vijay Shankar A$^{}$\thanks{E-mail: vijaynebula@gmail.com}
\\
$^{}$Center for Computational Engineering and Networking(CEN), Amrita School of Engineering, Coimbatore, Amrita Vishwa Vidyapeetam ,India\
}

\date{Accepted XXX. Received YYY; in original form ZZZ}

\pubyear{2020}

\begin{document}
\label{firstpage}
\pagerange{\pageref{firstpage}--\pageref{lastpage}}
\maketitle

\begin{abstract}
This work identifies the orientation of galaxies in the Galaxy zoo data set. The images are first identified by the number of principal components required to represent 99 percent of the  variance of the  image. K means clustering is used to separate the galaxies on the basis of their central brightness along with outlier separation. Spectral clustering is then used to separate circularly symmetric galaxies and the remaining   galaxies are identified according to their orientation as flat , left and right on the basis of the alignment of the main axis. It is also seen that spectral clustering fails to make this classification when the galaxy images are noisy and works only when applied on a smaller subset of the total number of images in the Galaxy zoo data set. This method also fails in the presence of multiple galaxies in the image, considering them as an individual entity. 

\end{abstract}


%%%%%%%%%%%%%%%%%%%%%%%%%%%%%%%%%%%%%%%%%%%%%%%%%%

%%%%%%%%%%%%%%%%% BODY OF PAPER %%%%%%%%%%%%%%%%%%

\section{Introduction}

Orientation of Galaxies is an important factor in the study of the galaxy formation process.Different models of cosmological structure evolution predict galactic orientation to be either random or follow a specific pattern within galactic clusters (\cite{g1999}).A parametric way  of measuring Galactic orientation is using the second-order momenta of the intensity distribution  \cite{stobie1980} as used in (\citet{Dario1992}. Simulations have been used to study Galactic alignment ( e.g. \citet{10.1093/mnras/sty2567}). Position angle catalogues have been used to study correlations associated with cosmological alignment of galaxies (e.g.\citet{10.1093/mnras/stx1977}.A detailed study on the importance of Galactic alignment in cosmology has been carried out in (\citet{Kirk_2015}. The study in this paper proposes a non parametric method to classify non circular symmetric  galaxies according to their alignment using unsupervised machine learning and dimensionality reduction techniques.Alignment in this paper refers to the axis passing through the galaxy and the image horizontal. Unsupervised learning techniques are a common tool these days to study galaxies, in fact a number of unsupervised machine learning techniques have been used to classify galaxies according to their morphology (e.g \cite{martin2019}.

A technique called Principal component analysis originally invented by Karl Pearson in 1901, see for example (\citet{savante1987}) is commonly used for dimensionality reduction which helps in visual inspection of clusters, a ubiquitous technique in cluster analysis. A common way of using the technique is to find the minimum number of principal components required to describe the variance of the image substantially (\citet{Callega2004}) where a conclusion is drawn on the number of principal components required to describe the entire set of galaxies under consideration. In this work I discriminate the data by the number of principal components required to describe 99 percent of the variance of the image, that is each individual image is marked by the number of principal components required to represent 99 percentage of the variance of the image and a set of images below a certain number of principal components is separated and taken for analysis.

K means (\citet{macqueen1967}) is a powerful clustering technique which identifies clusters according to the euclidean distance between the between the data points and each cluster is identified by the centroid which are iteratively varied until convergence. This algorithm has been used to identify spiral galaxies and elliptical galaxies (\citet{Barchi_2016}) in the galaxy images associated with the Galaxy zoo (\citet{10.1093/mnras/stt1458}). In this work I use K means to separate galaxies on their central brightness, in addition to outlier separation.In this context outlier refers to bright objects in addition to the galaxy under consideration in the image, which will interfere with further cluster analysis. 

I have so far illustrated the ground work to apply spectral clustering (\citet{Zhuzhunashvili_2017}) , which clusters the data on the basis of the eigen vectors of a graph Laplacian matrix (\citet{MERRIS1994143}. This algorithm is computationally efficient for a small number of clusters on a sparse data set \citet{6287457}. I use the K nearest neighbours graph \citet{DBLP:journals/corr/abs-0711-0189} with computational efficiency under consideration. The ground work has been laid in the sense that the spectral clustering algorithm works efficiently only on a much smaller number of images in the entire data set , hence  they are divided according to their principal components and the K means clusters.  I am able to classify the images of the Galaxy zoo (\citet{10.1093/mnras/stt1458}) into outlier , centrally bright and centrally dark. Further the circular symmetric galaxies are separated from centrally bright and centrally dark and the rest of the data is separated into flat , left and right, depending upon the alignment of the major axis of the Galaxy with the image horizontal.

\section{Methodology}

In this section I describe the procedure used in the project which includes description of the data set used , the method used for logical separation into smaller samples and the application of K means followed by the Spectral clustering step which produces the final result. 

\subsection{Dataset Description}
The data set used in this description is a part of the Galaxy zoo project (\citet{10.1093/mnras/stt1458}), this was used in  Kaggle Galaxy challenge. This was a competition for the best algorithm that can classify galaxies morphologically based on a label developed by a online volunteer program which classified galaxies according a dendrogram based on the Hubble classification scheme (\citet{10.1093/mnras/stt1458}) . The volunteers are asked to identify the galaxies' morphological parameters in a series of questions each associated with a particular morphological parameter. Cumulative probabilities with higher weights to questions marked higher in the order of questions. The more obvious and poignant aspects such as ellipticity or spirality of the galaxy, feature at the top or the questionnaire and more intricate questions like 'number of spiral arms' feature at the bottom, consequently, the uncertainties are also seen to increase mostly in a top to bottom manner.  So you have a label for this training set with probabilities for each morphological feature. This was a competition to find out the best algorithm , that would emulate the behaviour of the crowd in terms of classifying galaxies, with mean square error of the probabilities as the evaluation metric .There are a total of 61578 galaxy images each  424x424 coloured images, in the training set, which is used in this work.I follow a completely different classification methodology without considering this label.This work carries out a classification on the basis of symmetry and orientation rather than intricate morphology. Besides, the labels have considerable levels of uncertainty (\citet{10.1007/978-3-642-38610-7_14}) and thus there is a need to explore non visual classification techniques (e.g.\citet{martin2019}.
\subsection{Preprocessing}

The original 424x424 coloured images is converted into gray scale by using the weighed average of the coloured image. In this project we are only worried about shapes and orientation so the color of the image does not make any significant change. The grey scale 424x424 does not augur well with about 12 GB RAM provided by Google Colaboratory, where I run this project. Hence it has been resized to 69x69 using spline interpolation (\citet{1163154}. In both these cases I have used the Skiimage library \citet{scikit-image}. It can be easily verified through visual inspection that the properties of the galaxies under these changes are well retained for the purpose of this work. The images are then flattened into vectors and an array or list of those vectors is used as input. 

\subsection{PCA Based Separation}

\begin{figure}
 \includegraphics[width=\columnwidth]{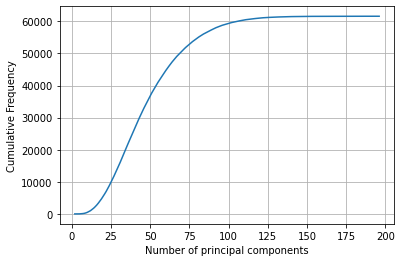}
 \caption{The cumulative plot of the number of principal components which are required to represent 99 percent of the variance of the image in the training images set of the Galaxy zoo data set}
 \label{fig:figure1}
\end{figure}

The Galaxy images are separated according to the number of principal components required to represent 99 percent of the variance of the original image.This is a technique commonly employed to decide the number of principal components to describe the image in a much reduced dimension space. 12 principal components described as 'eigengalaxies' are found to represent 96 percentage of the variance of the 30x30x3 cut outs of the galaxy images (\citet{Uzeirbegovic_2020}).A detailed description of the concept of explained variance ratio which is used as a metric to represent the variance of the given image using a certain number of principal can also be seen in (\citet{Uzeirbegovic_2020}). In this work, instead of fixing a number of principal components for all the images, each image is associated with the number of principal components required to represent 99 percent of the variance of the original image space.  The range extends from as little as two components to 196 components. The cumulative plot representing the number of principal components and the occurrence within the data set is given in figure \ref{fig:figure1}> . Images with components greater than 100, for example, are seen to have external objects like star to a greater degree than usual and the case of 'noisy galaxies' is also lesser.The inference is loosely analogous to the results in (\citet{Uzeirbegovic_2020}),where the morphology is seen to vary as a function of the Euclidean distance between the 'eigengalaxies'. Please note that the separation in this case is only weak  but nevertheless, I am able to provide a logical separation of the data set consisting of all the different types of galaxies  , yet having a limit on the number of principal components in common between them. This is an alternative to a naive random selection of galaxies in the data set. The need for smaller samples is necessitated by the constraints imposed by RAM. 

I work on two samples, the first with number of principal  components less than 26 consisting of 9799 images(sample1). The second set consists of images which have principal components from 26 to 35 inclusive, with 11030 images(sample2).In both cases the 69x69 grey scale image space is flattened out and we have an array of images with each row representing the flattened version of the image under consideration, say in sample1 there will be a array with dimension 9799x4761. One can see that in the second sample the number of 'noisy galaxies' are greater, as will be described in subsequent sections. By 'noisy galaxy' I mean the galaxies which have their own pixels spilling over to the background. This is another type of noise in addition to external objects. This type of noise is particularly challenging , and results in a galaxy of a very different shape being classified differently. This work has been carried out  on a smaller subset of the data divided into two parts as described above. But this can be easily be extrapolated to higher number of principal component galaxies , the biggest challenge will be that of the 'noisy galaxy'. 
\subsection{Kmeans Central Brightness Clustering}
\begin{figure}
 \includegraphics[width=\columnwidth]{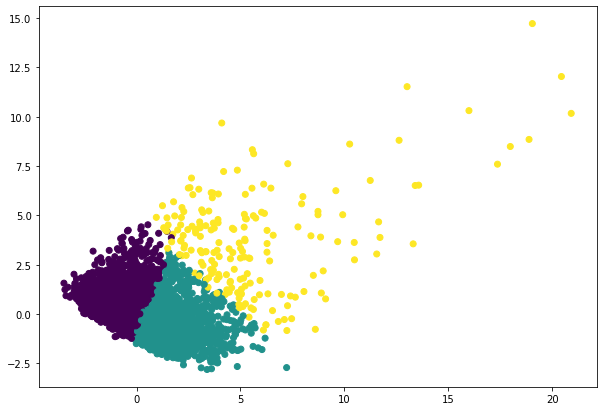}
 \caption{K means clusters of the galaxies which are represented by less than 26 principal components. violet represents the centrally dark galaxies,green centrally bright and the outliers are yellow. Notice how the outliers are looseley attached to each other.}
 \label{fig:figure2}
\end{figure}

This work uses K means algorithm to separate galaxies into three classes centrally bright, centrally dark and outlier classes. It is interesting to note that a correlation has been observed between eccentricity and brightness in the K means clusters of the galaxies \citet{Gauthier2016GalaxyMC}. The array of flattened vectors of the grey scale images used as input is reduced to two dimensions using Principal component analysis and are plotted against each other (cluster plot) and the K means clusters are indicated by the different colours for sample1 Fig. \ref{fig:figure2} . 
In this plot the dots represent the variance of the original set of images of sample1 in two dimensions. One can observe that the  cluster represented by yellow , with loosely packed dots represents the outlier , while the other two colours represent the centrally dark and centrally bright galaxies. It is also noteworthy that the K means cluster plot  for sample2 and other samples in the data set are exactly analogous to the plot in Fig.\ref{fig:figure2} , following a very similar pattern. This plot may epitomize the properties of the galaxy images to some extent. Note that central brightness refers to the overall brightness of the galaxy with respect to the center, outlier refers to the presence of bright external objects like stars. 

\subsection{Spectral Clustering}
This section elaborates the application of spectral clustering to the results of the K means clustering, centrally bright and centrally dark sets of images. The outlier class is ignored. First the samples under consideration are computed to obtain a K Nearest Neighbours square sparse matrix (\citet{10.1007/978-3-642-33260-9_22}) representing the interrelation between each individual pixel of the flattened images. The number of nearest neighbours is taken to be 30. There is no particular theory to estimate the number of nearest neighbors, through trial and error the number is fixed to be 30. 

\begin{figure}
 \includegraphics[width=\columnwidth]{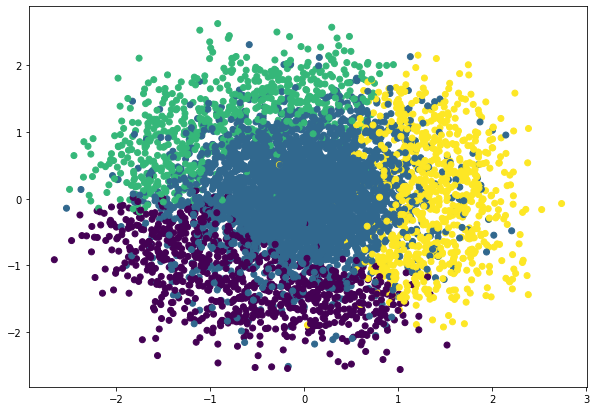}
 \caption{Spectral clusters of the galaxies which are represented by less than 26 principal components. Violet represents the flat galaxies,green circularly symmetric, yellow right aligned and blue is the left aligned galaxy .}
 \label{fig:figure3}
\end{figure}

\begin{figure}
 \includegraphics[width=\columnwidth]{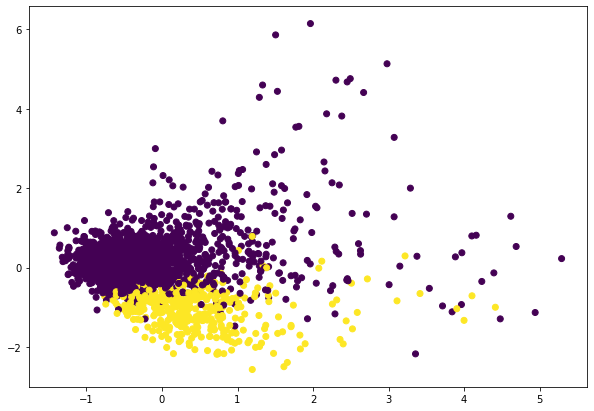}
 \caption{Spectral clustering applied to separate circular galaxies(violet) from flat  galaxies (yellow)}
 \label{fig:figure4}
\end{figure}

The PCA plot of clusters obtained for the galaxies with components less than 26(sample1)  and that are centrally dark are given in  the cluster plot in Fig.\ref{fig:figure3}. The galaxies are classified into flat , left, right and circularly symmetric on the basis of the overall alignment with respect to the horizontal axis of the image in the case of the former three and in the last case the circularly symmetric galaxies are taken as a separate class. Note that the flat galaxies are not perfectly flat with respect to the axis, but rather have alignments which are comparatively lesser to a great extent than that of the left and right aligned galaxies. A similar plot is obtained for centrally dark galaxies for images in sample2. In the case of centrally bright case of sample1 a further spectral clustering is carried out on the circular symmetric class and the flat and circular clusters are separated as in the cluster plot in Fig. \ref{fig:figure4} because the plot analogous to Fig. \ref{fig:figure3} does not produce good clustering results. 
In the case of centrally bright galaxies of sample 2  the separation between flat and circular galaxies is indiscernible. This is possibly due to the presence of 'noisy galaxies' where there is spillover of the galaxy pixels into nearby areas , masking the actual symmetry of the galaxy. One can note the correlation between centrally dark galaxies and spectral clustering giving good results.

\begin{figure}
 \includegraphics[width=\columnwidth]{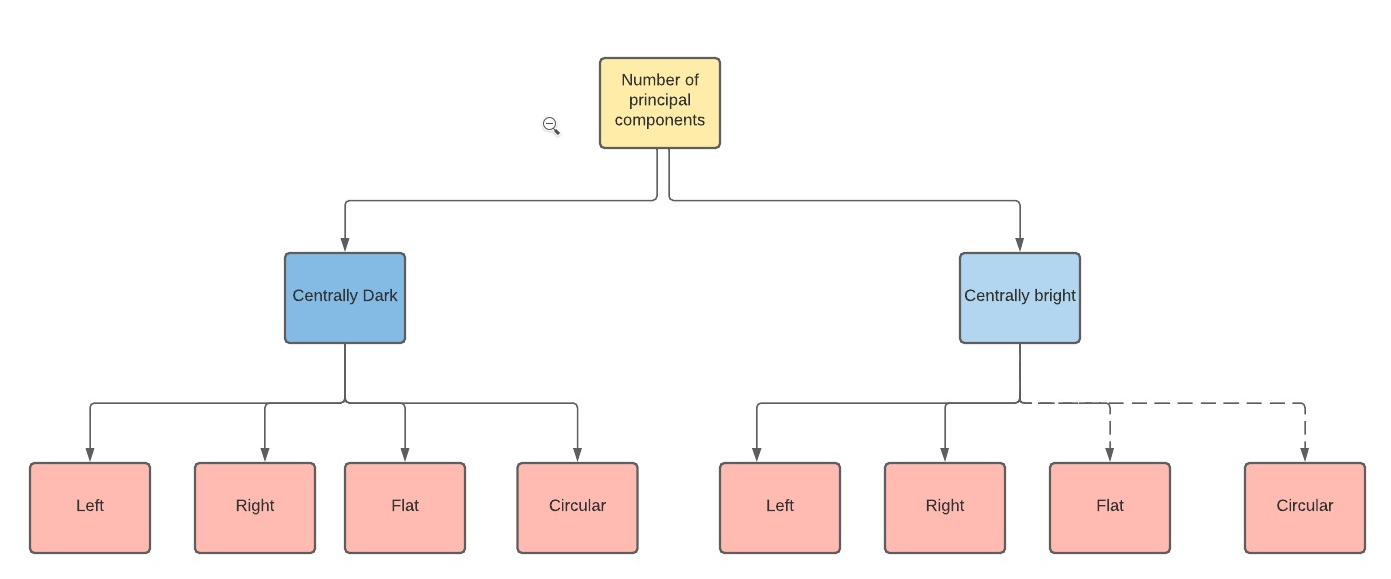}
 \caption{Block diagram explaining the methodology used in the work, The top block showing the first step of PCA based separation, The next block in blue indicating the K means separation and the last set of blocks displaying the the classes seperated by the spectral clustering. The dotted arrows indicate that the classification in those cases is not always guaranteed}
 \label{fig:blockdiagram}
\end{figure}

In all the above cases where I use  Principal Component Analysis, K Nearest Neighbours, K means and Spectral Clustering I use the Scikit-learn library \citet{scikit-learn}. A block diagram of the entire methodology is depicted in Fig. \ref{fig:blockdiagram} , where the difficulty in separating circular and flat galaxies in the case of centrally bright images is represented through a dotted line.

\section{Results And Discussion}

\begin{figure}%

    \subfloat{{\includegraphics[scale=.20]{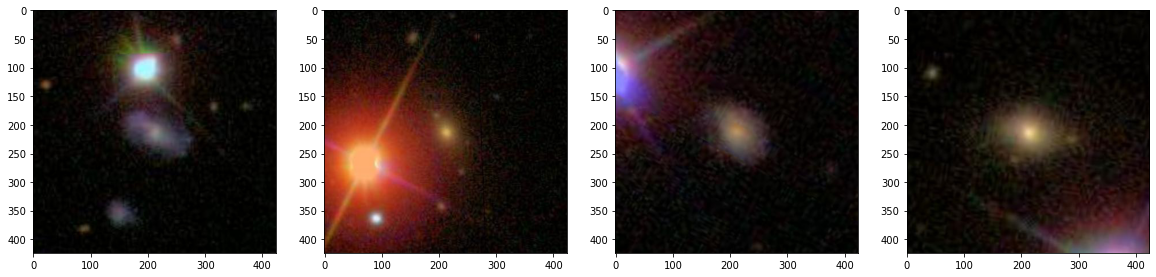} }}%

    \subfloat{{\includegraphics[scale=.20]{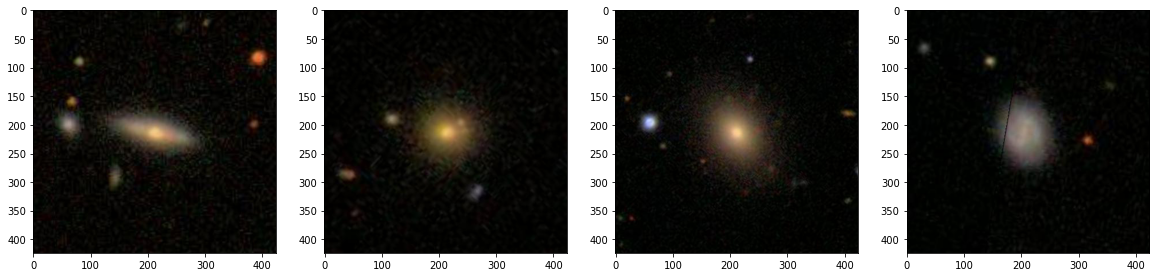} }}%

    \subfloat{{\includegraphics[scale=.20]{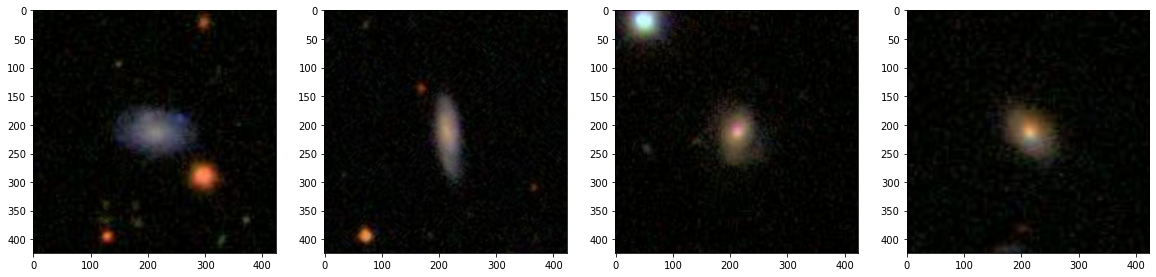} }}%

    \caption{The results of the K means clustering consisting of outlier class,centrally bright stars and centrally dark and centrally bright respectively, in rows, from top to bottom}%

    \label{figure5}%

\end{figure}

This paper makes a bold attempt at finding out the orientations of galaxies in the Galaxy Zoo data set training images. K means first classifies the galaxies into three classes, outliers, centrally dark and centrally bright as in Fig. \ref{figure5}, which shows the K means clustering of the sample2 images. Exactly Similar results can be obtained for sample 1. The images are grouped as outlier class at the top , with the presence of external stars, the second row from top centrally bright and the bottom row centrally dark.The last two rows are clustered according to their brightness , with focus on the center of the Galaxy. This includes , but is not limited to the Galactic Nucleus. There are cases where the distinction due to the brightness is quite subtle rather than prominent. However the galaxies have been cross verified several times through visual inspection to confirm that there is no other  more poignant intra-cluster correlation feature, than the central brightness. The effect is more prominently visible in case of galaxies which are yellow in colour rather than the galaxies which are blue or white in colour. The problem in the outlier class is that even the presence of a small bright object makes the image clustered as an outlier. Note that the results produced are that of galaxy images plotted in their 424x424 dimension in colour in contrast with the 69x69 flattened grey scale images used for computations.

\begin{figure}%

    \subfloat{{\includegraphics[scale=.20]{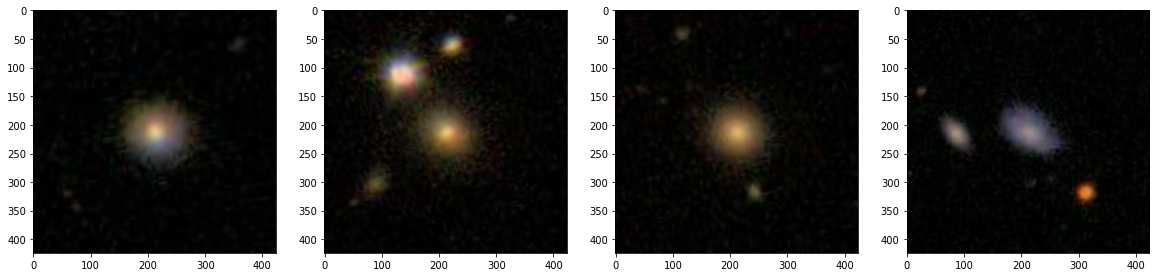} }}%

    \subfloat{{\includegraphics[scale=.20]{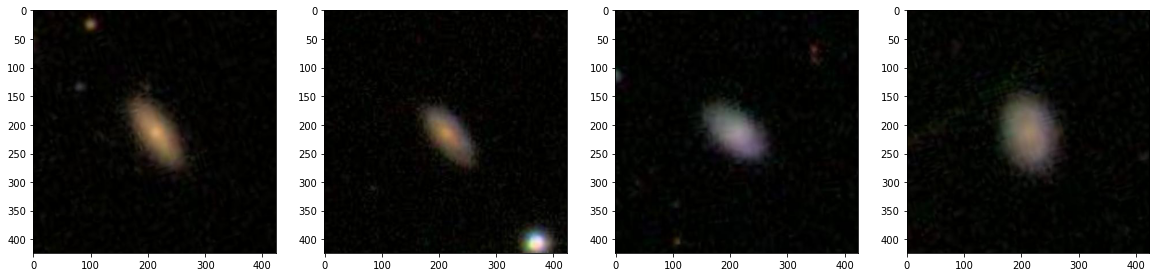} }}%

    \subfloat{{\includegraphics[scale=.20]{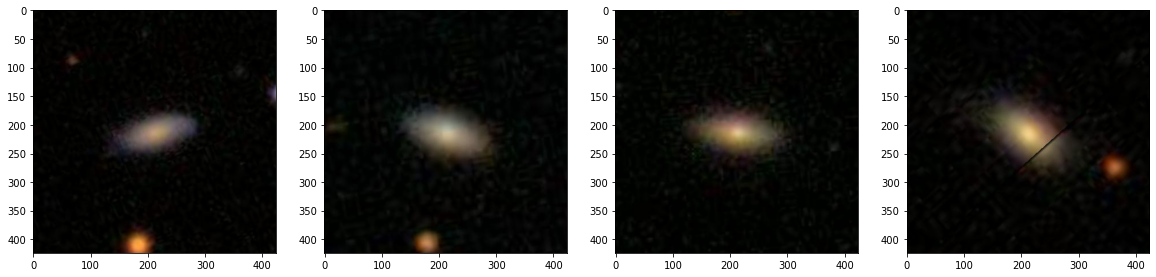} }}%
    
    \subfloat{{\includegraphics[scale=.20]{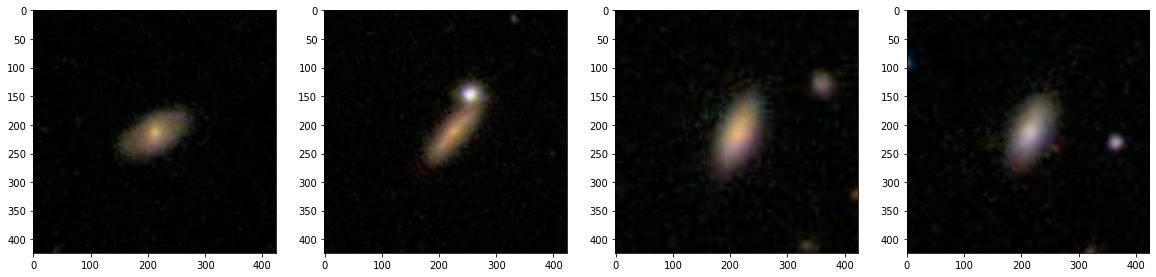} }}%

    \caption{The results of the Spectral Clustering on the centrally dark images of the sample1 showing four classes in order from top to bottom(row wise ) as Circularly Symmetric, Left , Flat and Right}%

    \label{figure6}%

\end{figure}

\begin{figure}%

    \subfloat{{\includegraphics[scale=.20]{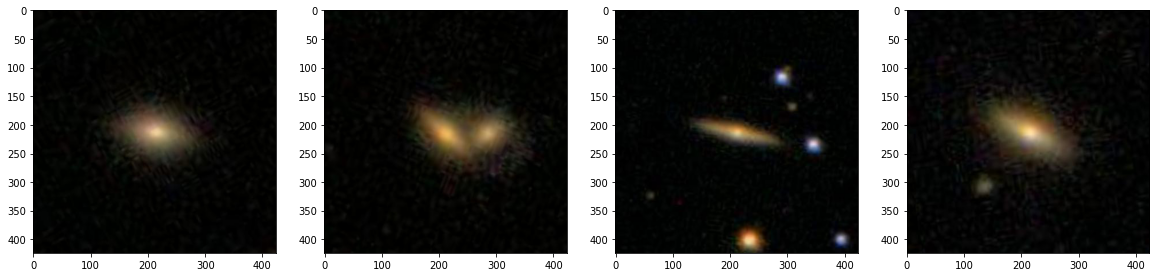} }}%

    \subfloat{{\includegraphics[scale=.20]{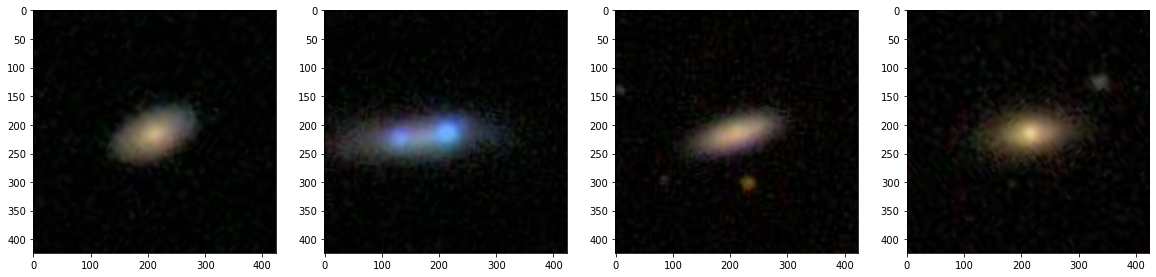} }}%

    \caption{The results of the Spectral Clustering on the 'flat' and 'centrally dark' images of the sample1 showing two classes left and right in order from top to bottom showing the sensitivity the spectral clustering technique has with respect to galactic orientation.}%

    \label{figure7}%

\end{figure}

In the case of Spectral clustering the results can be clearly seen in Fig.\ref{figure6} , this is effective only for the centrally dark images of sample 1, similar results can be seen in the case of sample 2 as well, for the centrally dark case. At this point one needs a clear definition of what is referred to as alignment or orientation in this paper. The alignment is the angle which the largest horizontal line passes through the Galaxy makes with the horizontal line passing through the center of the image. In the case of 'left' The galaxies make an angle  of about 120 degrees taken in the conventional anti clockwise direction. 'Right' Makes an angle of about 60 degrees with the horizontal. The 'Flat' class consists of galaxies which are tilted in both the directions but the angle is much smaller. Please note that the angles are only approximate and this work falls short of making any precise threshold measurement for any of these cases. In the case of the centrally bright case of sample 1, addition of spectral clustering once again to the case of the circular symmetric case results in the separation of the flat and the circularly symmetric galaxies as explained in Section 2.5. 
Besides that , there is an excellent proof of concept to how spectral clustering works in discerning orientations in galaxies . The flat galaxies contain both left and right aligned galaxies making a much smaller angle with the image horizontal. In  Fig. \ref{figure7} we can see further discernment into left and right aligned galaxies as separated by spectral clustering in the case of flat and  centrally dark case of sample 2. This result paves the door for the use of spectral clustering for identifying even small changes in the orientation of the galaxy. 

\subsection{Limitations and Future Work}
The biggest limitation of the work is that it does not make any reasonable predictions about Merger galaxies in the image separately. It only classifies them on the basis of the overall alignment. This is also true in the case of K means clustering where it is difficult to see which of the galaxies in the merger images is taken for consideration in the central brightness classification. The outlier class is merely ignored because of the presence of extremely bright stars which will obscure the analysis. 
The biggest limitation is the presence of 'noisy galaxies' which are images where galactic pixels spill over to neighbouring areas masking the actual symmetry.  This may easily fool both the K means and spectral clustering algorithms.
Besides, the spectral clustering for orientations works best on the centrally dark images and not as well on the centrally bright images.The biggest challenge and future work will be to overcome these inconsistencies possibly by separating undesirable elements and denoising the noisy galaxies. All these results have been verified visually through intensive visual inspection of randomly selected galaxies from the clustered results. There is a need for  statistical scrutiny or errors and misclassification, which will become easier once we are aware of the exact threshold angle which spectral clustering uses to make its orientation classification. Finding out the relation between the angle used in this work and the position angle commonly used in astronomy (\citet{1981csa..book.....T}) , will be an important task in the future. The work, although in its proof of concept stage, will pave the advent of a new paradigm in use of unsupervised learning in identifying orientations of galaxies.

\section{Conclusions}

This work acts as a good proof of concept in the application of an unsupervised machine learning technique to sort galaxies according to their orientation. In addition sufficiently good results have been obtained in separation of images with extremely bright stars. It is also shown how much spectral clustering is sensitive to even minor changes in the alignment of the galaxies.The ability of spectral clustering in separating circular symmetric galaxies is also evident. All this makes us conclude that spectral clustering is a very good tool in identifying orientation and symmetry of galaxies in a sky survey. The work has its limitations which include the challenges faced due to 'noisy galaxies'(pixel spillover), inability to say anything about merger galaxies and the omission of images with bright additional objects. The results are not completely precise but the overall results are noteworthy and will play a significant role in finding the orientation of galaxies in sky surveys. 

\section*{Data Availability}

The data used in this work is the one associated with the Galaxy zoo kaggle competition which can be accessed from \url{https://www.kaggle.com/c/galaxy-zoo-the-galaxy-challenge/data}. Please use the training images associated with the data set. I have included the list of images in the Git-hub repository which are a set of address strings to access Galaxy zoo training set. \url{https://github.com/tensorvijay/Galaxy_orientatation}. All of those collections have been given appropriate names relating to the work in this paper. Please refer to the 'README' section in the above link for further details about how to access the data set and verify the work presented in this paper.



\bibliographystyle{mnras}
\bibliography{example} % if your bibtex file is called example.bib

%\begin{thebibliography}{99}
%\end{thebibliography}

%%%%%%%%%%%%%%%%%%%%%%%%%%%%%%%%%%%%%%%%%%%%%%%%%%

%%%%%%%%%%%%%%%%% APPENDICES %%%%%%%%%%%%%%%%%%%%%

%%%%%%%%%%%%%%%%%%%%%%%%%%%%%%%%%%%%%%%%%%%%%%%%%%

% Don't change these lines
\bsp	% typesetting comment
\label{lastpage}
\end{document}